Envelope Soliton Oscillator for UWB

Gary J. Ballantyne, February 2008

An electronic oscillator is shown to support an envelope soliton pulse. The oscillator comprises a loop, formed by a nonlinear transmission line, an amplifier, a bandpass filter and a saturable absorber. The soliton is suitable for ultra-wideband radio-frequency communication; its existence and stability are demonstrated with simulations.

*Introduction:* A soliton is a naturally occurring solution to certain famous nonlinear wave equations [1]. Broadly, solitons are pulses that survive interaction with other solitons, and thus behave somewhat like a particle. An oscillator with baseband soliton modes, the Baseband Soliton Oscillator (BSO), was reported in [2,3]. A similar envelope soliton oscillator (ESO) was also proposed but has not been published (outside of [4]). Recent publications have considered baseband electrical solitons in integrated circuits [5], and even an integrated realization of the BSO [6]. Because of this interest, and the potential for application to developing ultra-wideband (UWB) systems [7], this letter considers the basic properties of the ESO.

UWB systems employ short pulses that occupy a broad spectrum. Gaussian pulses, or variations thereof, are often considered for their time-bandwidth properties. An interesting property of the Gaussian envelope is that its Fourier transform is also Gaussian. The ESO soliton is similar — it shares the self-transform property — and also has a low time-bandwidth product. Thus, further to the interesting dynamics of the ESO, it is potentially a source of UWB pulses.

Similar to related optical devices [8], the ESO essentially comprises a nonlinear transmission line (NLTL) formed into a loop with an amplifier to compensate for losses. Our goal is to show that an envelope soliton is a stable fundamental mode of this system. That is, that a soliton forms and circulates endlessly in the loop. Though more work is required to realize a practical device it is essential to first prove this fundamental premise.

*Soliton Equation*: We consider propagation of the complex variable $u$, where the physical voltage is $v = u + u^*$. Then $u = u(X,t)$ can be described by the nonlinear Schrödinger (NLS) equation with perturbation $R$ [9]

$$i\frac{\partial u}{\partial t} + P\frac{\partial^2 u}{\partial X^2} + Q|u|^2 u = R(u). \qquad (1)$$

Here $t$ is time, and $X$ is a spatial variable in a frame of reference traveling at the group velocity; $P$ and $Q$ are coefficients of dispersion and nonlinearity, and $R$ includes all other effects (attenuation, gain, filter, etc). With $R = 0$, equation (1) has a pulse solution, with amplitude $A$, given by

$$u = A\operatorname{sech}\left(A\sqrt{Q/2P}\,X\right)\exp\left(iA^2 Qt/2\right). \qquad (2)$$

Equation (2) remains a good approximate solution if the timescales of $R \neq 0$ are much shorter than those of the dispersion and nonlinearity. This is known from average-soliton theory [10,11], where the pulse experiences a (possibly large) periodic change in amplitude — in this case from attenuation in the NLTL and compensating amplification $\mu$. The upshot is that the nonlinear coefficient, $Q$, is multiplied by $2\ln\mu/1-\mu^{-2}$.

*Oscillator Description*: To support a pulse $P$ and $Q$ must be such that $PQ > 0$, which is possible with various discrete, and semi-discrete NLTL [5,9]. We consider a straightforward example with the usual series-inductor ($L_0$) / shunt-varactor ($C(v)$) ladder topology.

Though envelope solitons can propagate in lines with common PN junction varactors, for simplicity we assume a nonlinear capacitance approximated by $C(v) = C_0(1-3\phi v^2)$. Here $C_0$ is the nominal capacitance, and $\phi$ determines the nonlinearity (e.g. see the MOSVAR characteristic in [5]). The coefficient of nonlinearity is then $Q = 3\phi\omega/2$, and the dispersion is $P = -\omega/8$, where $\omega$ is the centre frequency of the pulse.

Figure 1 shows the ESO. In addition to the NLTL and amplifier the oscillator includes a bandpass filter (BPF) and saturable absorber (SA). For simplicity the BPF is formed from a series resistor ($R_F$) followed by a shunt inductor-capacitor resonator ($L_F, C_F$). The BPF helps choose the centre frequency of the soliton, and also stabilizes the device. To understand the stability mechanism we note that an increase in soliton amplitude increases its bandwidth and causes a greater loss through the BPF (and visa versa).

With periodic boundary conditions equation (1) has many interesting solutions, including cnoidal multiple-pulse waveforms. From simulations it is apparent that the SA selects a single-pulse, suppressing the formation of smaller pulses [4]. For demonstration we have assumed the SA to be given by $y = x/(1+(0.01/x)^4)$. This characteristic prevents the ESO from self-starting from noise, though, the SA circuit in [6] addresses this issue.

*Simulation*: Figure 2 shows two simulated waveforms at the NLTL input (heavy lines) corresponding to gains of (a), $\mu = 2.02$, and (b), $\mu = 2.04$. The parameter values are $L_0 = 1$ mH, $C_0 = 110$ pF, $\phi = 0.09$ V$^{-2}$, $R_F = 1\,\Omega$, $C_F = 131$ nF, and $L_F = 1\,\mu$H. The transmission line is 75 sections long, with an additional 5 sections to minimize reflections at the input of the amplifier. The line is ultimately terminated with a matched load, assuming oscillation at the BPF centre frequency, with a resistance of $0.89\sqrt{L_0/C_0}$ [12]. Losses in the line are caused by a 50 $\Omega$ resistor in series with each transmission line inductor and 3 $\Omega$ in series with each capacitor. These parameter values are consistent with [4]; though the oscillation frequency is far too low for UWB, the simulation can be trivially scaled

(via $L_0$ and $C_0$). To emphasize this point, the time units in figure 1 are scaled according to $t = \sqrt{L_0 C_0}\,\tau$ .

The lighter lines in figure 2 show the envelope of equation (2), allowing for the transformation from the complex envelope, and average-soliton theory. The characteristic soliton relationship between amplitude and width is evident. Both waveforms evolved from an initial condition of 1V on the first transmission line capacitance. Naturally, we cannot simulate every possible initial condition, or for infinite time, but our simulations compellingly imply this mode is an attractor, and that it is stable.

*Discussion*: We have demonstrated that an envelope soliton is a stable mode of a simple electrical oscillator. Clearly much work is required to evolve this work into a practical device, and the technological challenges should not be understated. Indeed, many important questions — such as robustness and the best system parameters — would be better answered experimentally. Nevertheless, the existence of a stable soliton mode is necessarily the first and most important, step.

Solitons have properties that may prove useful for UWB. The pulse bandwidth is controllable through the loop gain, and its centre frequency via the BPF. Although the best modulation method and fit with the UWB spectral mask needs exploration, the device is a promising source of electrical solitons.

Solitons have a rich heritage of theoretical analysis and experimental confirmation — especially in the field of optical communications. Indeed, communication systems based on the special properties of solitons have been proposed [13]. Whether these advanced topics can be applied to UWB is speculative, but intriguing nonetheless.

Captions

Figure 1: Envelope soliton oscillator

Figure 2: Waveforms at NLTL input (heavy lines) and NLS envelope for (a), $\mu = 2.02$, and (b), $\mu = 2.04$.

Figure 1

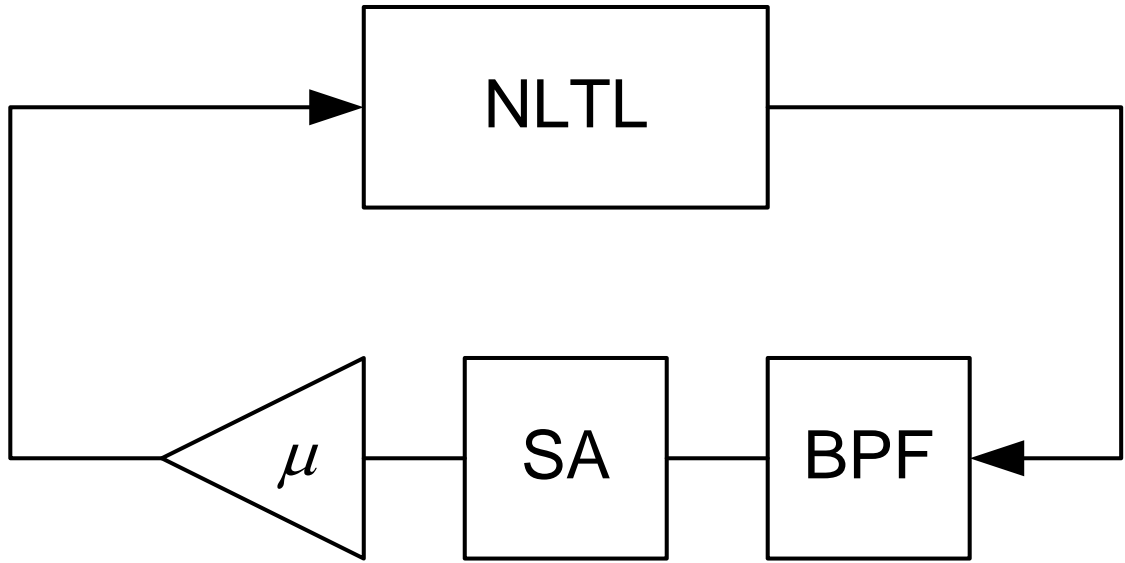

Figure 2

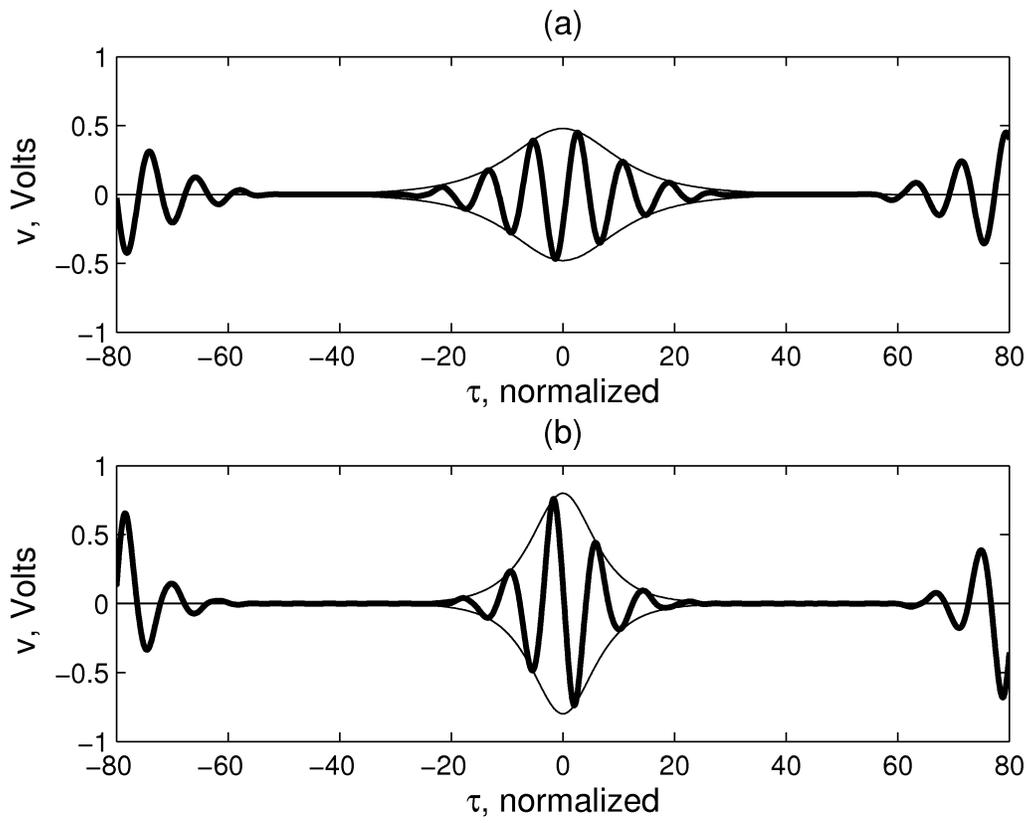